\documentclass[a4paper,12pt]{article}
\usepackage{amsmath, amssymb, amsthm}
\usepackage{fullpage}
\usepackage{hyperref}
\usepackage{graphicx}
\usepackage{parskip}
\usepackage{color}
\usepackage{appendix}

\let\originalleft\left
\let\originalright\right
\def\left#1{\mathopen{}\originalleft#1}
\def\right#1{\originalright#1\mathclose{}}

\numberwithin{equation}{section}

\begin{document}
\title{Hyperbolic Skyrmions}
\author{Thomas Winyard\\[10pt]
{\em \normalsize Department of Mathematical Sciences, }\\{\em \normalsize Durham University, Durham, DH1 3LE, U.K.}\\[10pt]
{\normalsize t.s.winyard@durham.ac.uk}}

\maketitle
\vspace{15pt}
\begin{abstract}
We investigate $SU(2)$ Skyrmions in hyperbolic space, by computing numerical solutions of the nonlinear field equation. We first demonstrate the link between increasing curvature and the accuracy of the rational map approximation to the minimal energy static solutions. We investigate the link between Skyrmions with massive pions in Euclidean space and the massless case in hyperbolic space, by relating curvature to the pion mass. Crystal chunks are found to be the minimal energy solution for increased curvature as well as increased mass of the model. The dynamics of the hyperbolic model are also simulated, with the similarities and differences to the Euclidean model noted. 
\end{abstract}

\pagebreak

\section{Introduction}
The Skyrme model \cite{Skyrme:1961vq} is a (3+1)-dimensional nonlinear theory of pions that admits topological soliton solutions, called Skyrmions, which represent baryons. This has been well studied \cite{bible} with solutions calculated for a large range of topological charges \cite{Battye:2001qn}. The addition of a mass term has little effect on solutions of low baryon number, which continue to form shell like structures. However for larger charge solutions, a mass term starts to favour minimal energy solutions formed of finite chunks of a Skyrme crystal \cite{Castillejo1989801,crystalchunk,PhysRevD.87.085034}. 

It has been demonstrated that there is a surprising similarity between Skyrmions with massive pions in Euclidean space and the massless case in hyperbolic space \cite{Atiyah2005106}. The cited paper also outlines a method for constructing Skyrmions with massive pions from instanton holonomies, by first modelling a hyperbolic Skyrmion by taking holonomies along particular circles in $\mathbb{R}^4$ \cite{0305-4470-23-16-022} and applying a mapping relating hyperbolic curvature and Euclidean mass to produce the Euclidean Skyrmion \cite{Atiyah2005106}.

This posits that there could be a geometrical underpinning to the standard mass term, traditionally used in the Skyrme model. This suggests that understanding Skyrmions in hyperbolic space and the affect that curvature has, may shed some light on Skyrmions with massive pions in Euclidean space. Most notably there are certain properties for Skyrmion solutions in Euclidean space, that only occur once the mass term is turned on, or exceeds a certain threshold. Namely, the formation of crystal chunk solutions, as the global minima, for higher charge systems that exceed the threshold mass. If some similar behaviour were to be observed for massless solutions in hyperbolic space, it would support this geometric link. In fact, it will be demonstrated that the map linking the curvature of hyperbolic Skyrmions with massive Euclidean solutions, can be used to predict the global minima solution. 

We will also examine the dynamics of Skyrmions in hyperbolic space, demonstrating that they scatter along geodesics, with maximally attractive channels corresponding to a relative rotation through an angle $\pi$, about an axis orthogonal to the connecting geodesic.

\section{The Model}
The Lagrangian density for an $SU(2)$ valued Skyrme field $U(t,\boldsymbol{x})$ is given by,
\begin{equation}
\mathcal{L} = -\frac{1}{2} Tr\left(R_\mu R^\mu\right) + \frac{1}{16} Tr\left(\left[R_\mu,R_\nu\right]\left[R^\mu,R^\nu\right]\right) - m^2_\pi Tr\left(U - 1_2\right)
\label{lagden}
\end{equation}
where $R_i = \left(\partial_i U\right) U^\dagger$ is the right $su\left(2\right)$ valued current. The associated energy for a static Skyrme field $U(\boldsymbol{x})$ defined on a general Riemannian manifold $M$ with metric $ds^2 = g_{ij}dx^idx^j$ is

\begin{equation}
 E = \frac{1}{12\pi^2} \int \left\{ -\frac{1}{2}Tr\left(R_i R^i\right) - \frac{1}{16}Tr\left(\left[R_i,R_j\right]\left[R^i,R^j\right]\right) + m^2 Tr\left(1-U\right)\right\} \sqrt{g} d^3 x
 \label{energy}
\end{equation}

where $g$ is the determinant of the metric. Note that both of the above expressions have the parameters preceding the first two terms scaled out. $m$ is the tree-level mass of the pions, which can be observed by using the $SU\left(2\right)$ nature of the field and writing the equation in terms of pion fields $U = \sigma + i\boldsymbol{\pi}\cdot\boldsymbol{\tau}$, where $\boldsymbol{\tau}$ is the triplet of Pauli matrices and $\boldsymbol{\pi} = \left(\pi_1, \pi_2, \pi_3\right)$ the triplet of pion fields. 

Much work has been done on the solutions to this equation for Euclidean space $M = \mathbb{R}^3$ upto topological charge $108$ \cite{Battye:2001qn,PhysRevD.87.085034}. However we are interested in considering Skyrmion solutions in hyperbolic 3-space $M = \mathbb{H}^3_\kappa$, which is the space with constant negative curvature $-\kappa^2$. The metric of $\mathbb{H}^3_\kappa$ takes the form, 

\begin{equation}
 ds^2 \left(\mathbb{H}_\kappa^3\right) = d\rho^2 + \frac{\sinh^2\left(\kappa \rho\right)}{\kappa^2}\left(d\theta^2 + \sin^2\theta d\phi^2\right).
\end{equation}

where $\rho$ is the hyperbolic radius. If we take the limit of zero curvature, we recover the Euclidean metric, with the hyperbolic radius equal to the standard Euclidean radius $\rho = r$. We will also make use of the standard Poincare ball model for displaying results. This can be obtained from the above metric by a simple radial transformation $\rho = \frac{2 \tanh^{-1}{\left(\kappa R\right)}}{\kappa}$, to give the following metric, 

\begin{equation}
 ds^2 \left(\mathbb{H}_\kappa^3\right) = \frac{4\left(dR^2 + R^2\left(d\theta^2 + \sin^2\theta d\phi^2 \right) \right)}{\left(1-\kappa^2 R^2\right)^2}.
\end{equation}

Hence our space can be modelled by a sphere with a boundary at infinite hyperbolic radius given by $R=\frac{1}{\kappa}$ (though our plots will always be scaled to an equivalent size).

The vacuum for the massless theory is any constant $U$, however the inclusion of the mass term $m>0$ gives the unique vacuum to be $U = 1_2$. We will impose the boundary condition $U \rightarrow 1_2$ as $\rho \rightarrow \infty$, which is required for finite energy. This gives us a map $U:\mathbb{H}^3_\kappa\cup \left\{\infty\right\} = S^3 \rightarrow S^3$, and hence a topological charge as an element of the 3rd homotopy group, equivalent to an integer $B \in \pi_3\left(S^3\right) = \mathbb{Z}$,

\begin{equation} 
B = -\frac{1}{24\pi^2}\int \epsilon_{ijk}Tr\left(R_i R_j R_k\right) d^3 x.
\end{equation}

\section{Approximations}
There are a few approximations for Skyrmions with massless and massive pions. The rational map approach will be the most useful in this paper. The angular dependence of the solution is approximated to be a rational map between Riemann spheres \cite{Houghton:1997kg}. On extension to massive pion solutions, it is found that only shell-like approximations can be closely approximated. While multi-shell like solutions have been modelled in an attempt to form more crystal like solutions \cite{Manton:2000kj}, they are poor approximations to the full minimal energy solutions. They can be useful for initial conditions in numerical simulations however. 
 
\subsection{B=1}
\begin{figure}
\begin{center}
\begin{tabular}{c c c}
\includegraphics[scale=0.33,natwidth=1000,natheight=1000]{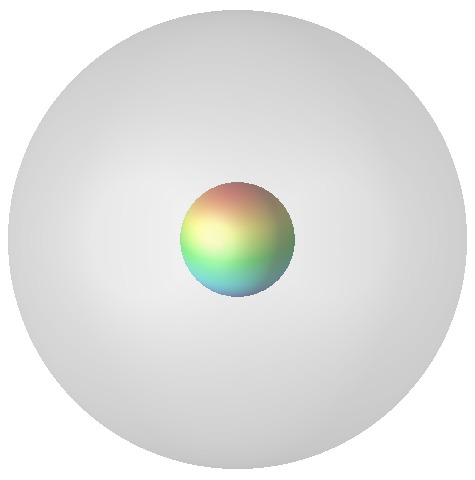} & \includegraphics[scale=0.33,natwidth=1000,natheight=1000]{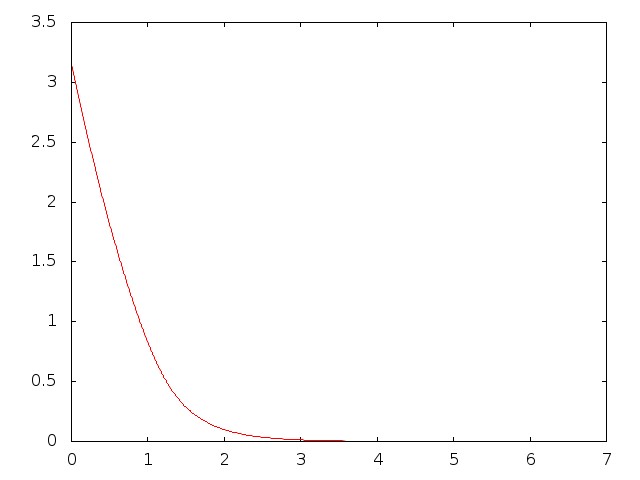} & \includegraphics[scale=0.33,natwidth=1000,natheight=1000]{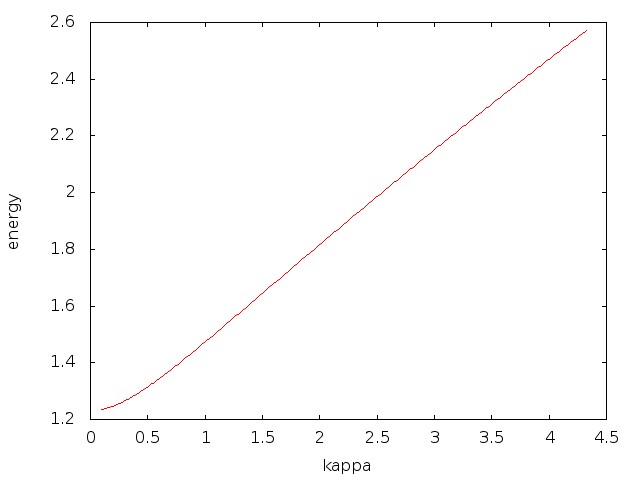}\\
(a) energydensity plot & (b) profile function $f\left(\rho\right)$  & (c) energy for increasing curvature\\
 & ($\kappa = 1$,$m=0$) & ($m=0$)
\end{tabular}
\end{center}
\caption{$B=1$ static hedgehog solution, (a) energy density plot in Poincar\'e ball, where grey shaded region represents the boundary of hyperbolic space, (b) profile function $f\left(\rho\right)$ for $\kappa = 1$, $m=0$, (c) energy for increasing curvature, for $m=0$.}
\label{B1static}
\end{figure}

In Euclidean and hyperbolic space the single Skyrmion solution can be  reduced to solving an ODE, using the hedgehog ansatz. This is known as a hedgehog solution due to its radial nature, as can be seen in figure \ref{B1static}. The field is given to be

\begin{equation}
 U = \exp\left(i f\left(\rho\right) \boldsymbol{\hat{x}} \cdot \boldsymbol{\tau}\right),
\end{equation}

where $\hat{\boldsymbol{x}} = \left(\sin{\theta}\cos{\phi},\sin{\theta}\sin{\phi},\cos{\theta}\right)$ is the unit vector in Cartesian coordinates, $f\left(\rho\right)$ is a monotonically decreasing radial profile function with boundary conditions $f\left(0\right) = \pi$ and $f\left(\infty\right) = 0$. Substituting this into the energy in (\ref{energy}) we get a radial energy of the form,

\begin{equation}
 E = \frac{1}{3\pi} \int \left( f'^2 \frac{\sinh^2\kappa\rho}{\kappa^2} + 2\left( f'^2 + 1 \right)\sin^2 f + \frac{\kappa^2 \sin^4 f}{\sinh^2 \kappa \rho} + 2m^2 \frac{\sinh^2\kappa \rho}{\kappa^2}\left(1 - \cos f\right)\right) d\rho
 \label{rationalenB1}
\end{equation}

The profile function $f\left(\rho\right)$ can then be found by minimising the above energy and is also shown in figure \ref{B1static} for $\kappa = 1$,$m=0$. This yields a function with an exponential asymptotic decay for $m=0$,

\begin{equation}
 f \sim A e^{-2\kappa \rho}.
\end{equation}

This takes a similar form to that of massive Euclidean Skyrmions ($\kappa = 0$) $f \sim \frac{A}{r}e^{-mr}$, but dependent on the curvature rather than the mass of the theory. This suggest a relation between curvature and mass. In fact it is found that if you select the correct curvature, you can produce an extremely similar profile function for any Skyrmion with massive pions in Euclidean space. See \cite{Atiyah2005106} to observe the graph showing the relation between $\kappa$ and $m$.

\subsection{Shell-like multisolitons}
Shell-like solutions can be well approximated by the rational map ansatz. In hyperbolic space this takes the following form,

\begin{equation}
 U\left(\rho,z\right) = exp\left[ \frac{i f\left(\rho\right)}{1 + \left| R \right|^2}\left( \begin{array}{c c} 1-\left| R \right|^2 & 2\bar{R} \\ 2R & \left| R\right|^2 - 1 \end{array} \right) \right]
 \label{rationalmap}
\end{equation}

where $z = e^{i\phi}\tan\left(\frac{\theta}{2}\right)$ is the Riemann sphere coordinate and $R\left(z\right)$ is a degree $B$ rational map between Riemann spheres. Substituting this ansatz into (\ref{energy}) we get the following radial energy,

\begin{equation}
 E = \frac{1}{3\pi}\int \left(f'^2 \frac{\sinh^2\left( \kappa \rho\right)}{\kappa^2} + 2B\left(f'^2 + 1\right)\sin^2f + \mathcal{I}\frac{\kappa^2 \sin^4 f}{\sinh^2 \left( \kappa \rho\right)} + 2 m^2 \frac{\sinh^2\left(\kappa \rho\right)}{\kappa^2}\left(1 - \cos f\right)\right)d\rho,
 \label{multirationalen}
\end{equation}

where

\begin{equation}
 \mathcal{I} = \frac{1}{4\pi}\int \left( \frac{1 + \left|z\right|^2}{1 + \left|R\right|^2}\left|\frac{dR}{dz}\right|\right)^4 \frac{2i dz d\bar{z}}{\left(1 + \left|z\right|^2\right)^2}.
\end{equation}

$\mathcal{I}$ is an integral to be minimised by the choice of rational map $R(z)$. Note that $\mathcal{I}$ is independent of $\kappa$ and hence the values match those in Euclidean space. The minimal values of $\mathcal{I}$ and the associated rational maps can be found in \cite{Battye:2001qn} for a range of values of $B$. Note that the earlier hedgehog ansatz is recovered for $B=1$, where $R = z$ is the minimising map, with $\mathcal{I} = 1$ and (\ref{multirationalen}) reduces to (\ref{rationalenB1}). 

This approximation will be used in various way to form initial conditions for the numerical computations presented later. We will also investigate how curvature affects the accuracy of the approximation.
 
\section{Static Solutions}
\subsection{Shell-like Static Solutions}
The static equations that follow from the variation of (\ref{energy}) were solved using a time dependent $4$th-order Runga-Kutta method to evolve the time-dependant equations of motion that follow from the relativistic lagrangian (\ref{lagden}), cutting the kinetic energy whenever the potential increased. The grid was modelled using the Poincar\'e ball model of radius $\kappa^{-1}$ on a cubic grid with $(201)^3$ grid points and lattice spacing (for the standard $\kappa = 1$) $\Delta x=0.005$. Spatial derivatives have been approximated using a $4$th-order finite difference method. We must fix the boundary at $R=\kappa^{-1}$ to be the vacuum at spatial infinity $U_\infty = \boldsymbol{1}_2$, to ensure finite energy. For all our simulations the topological charge, when computed numerically, gives an integer value to five significant figures, indicating the accuracy of the results.

Two forms of initial condition were considered. The rational map ansatz shown in (\ref{rationalmap}) and the product ansatz $U\left(\boldsymbol{x}\right) = U_1\left(\boldsymbol{x}\right) U_2\left(\boldsymbol{x}\right)$, which was used to place lower charge solitons at various well separated positions about the grid.

\begin{figure}
\begin{center}
\begin{tabular}{c c c}
\includegraphics[scale=0.4,natwidth=1000,natheight=1000]{B1.jpeg} & \includegraphics[scale=0.4,natwidth=1000,natheight=1000]{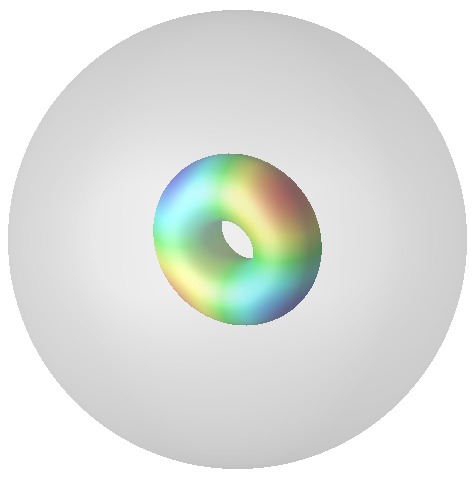} & \includegraphics[scale=0.4,natwidth=1000,natheight=1000]{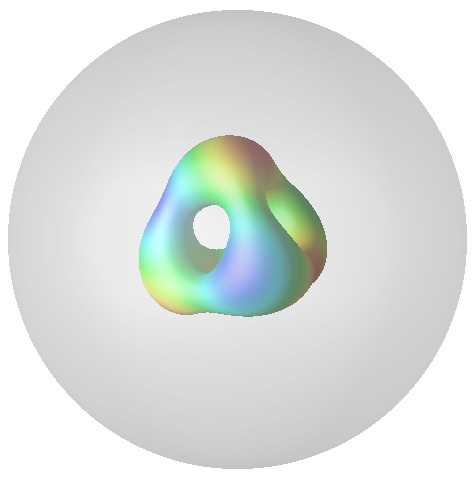}\\
(a) $B=1$ & (b) $B=2$ & (c) $B=3$\\
\includegraphics[scale=0.4,natwidth=1000,natheight=1000]{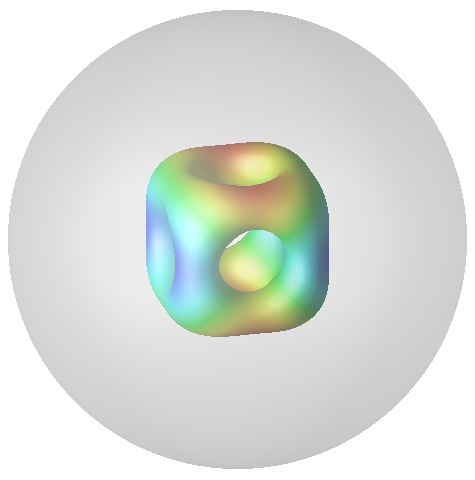} & \includegraphics[scale=0.4,natwidth=1000,natheight=1000]{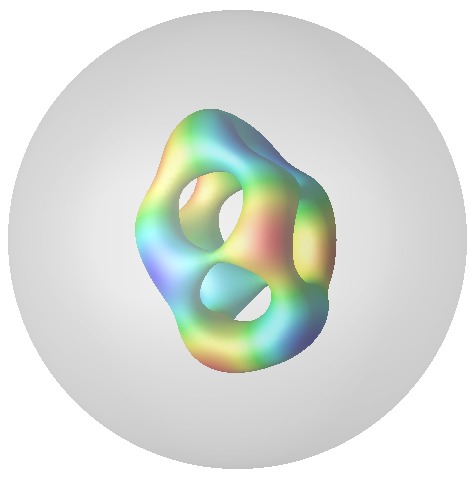} & \includegraphics[scale=0.4,natwidth=1000,natheight=1000]{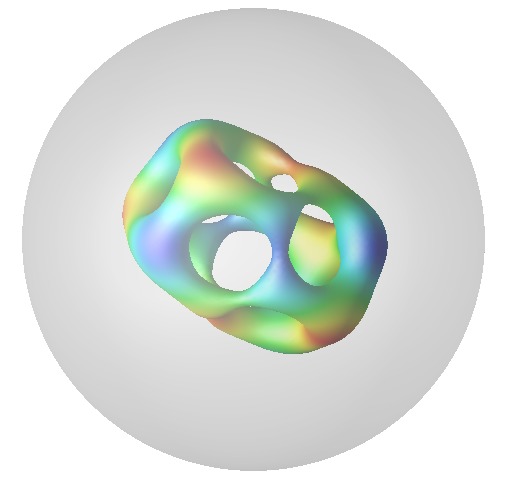}\\
(d) $B=4$ & (e) $B=5$ & (f) $B=6$
\end{tabular}
\begin{tabular}{c c}
\includegraphics[scale=0.4,natwidth=1000,natheight=1000]{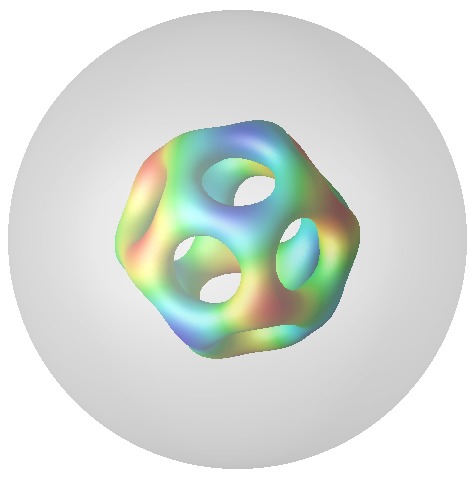} & \includegraphics[scale=0.4,natwidth=1000,natheight=1000]{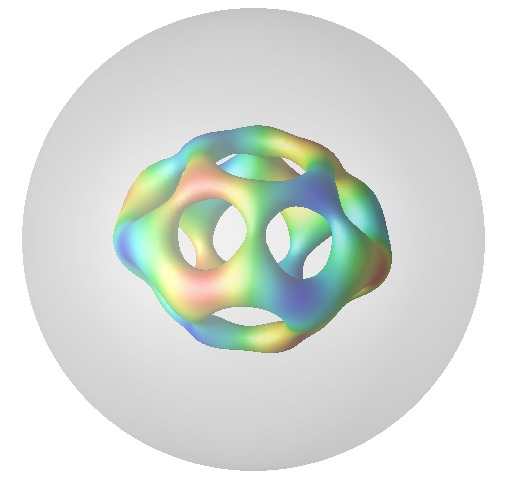}\\
(g) $B=7$ & (h) $B=8$
\end{tabular}
\caption{Energy isosurfaces of the shell like solutions with $\kappa = 1,m=0$ for $B=1-8$. The images are coloured based on the value of $\pi_2$ and the grey sphere represents the boundary of space in the Poincare ball model.}
\label{standardsolutions}
\end{center}
\end{figure}

The first eight shell-like static solutions for $\kappa = 1, m=0 $ can be seen in figure \ref{standardsolutions}. These solutions take a similar form to the Euclidean solutions of the same charge, with a few subtle differences. The faces of the polyhedron now appear to take the form of geodesic surfaces (a surface that contains curves belonging to the set of geodesics within the global space). Additionally, translating the solutions about the grid alters the apparent shape and means that lines of symmetry fall along geodesics of the space. This can be seen in more detail in the analysis of the $B=8$ solution in figure \ref{B8}. The crystal chunk solution clearly demonstrates a bowing of the line connecting the two $B=4$ solitons, this line is found to be a geodesic of the space.  

\begin{figure}
\begin{center}
 \includegraphics[scale=0.6,natwidth=1000,natheight=1000]{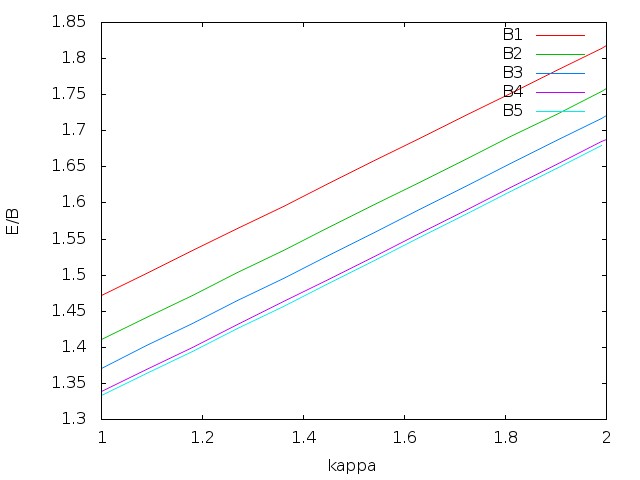}
\end{center}
\caption{A plot of the energy for charge $B=1-5$ shell like solutions against $\kappa$}
\label{curvyenergy}
\end{figure}

If we look at the energies displayed in table \ref{energies} we can see the expected trend in energies for increasing charge. We also observe how the energy of a given charge solution scales with curvature in figure \ref{curvyenergy}. 

\begin{figure}
\begin{center}
 \includegraphics[scale=0.6,natwidth=1000,natheight=1000]{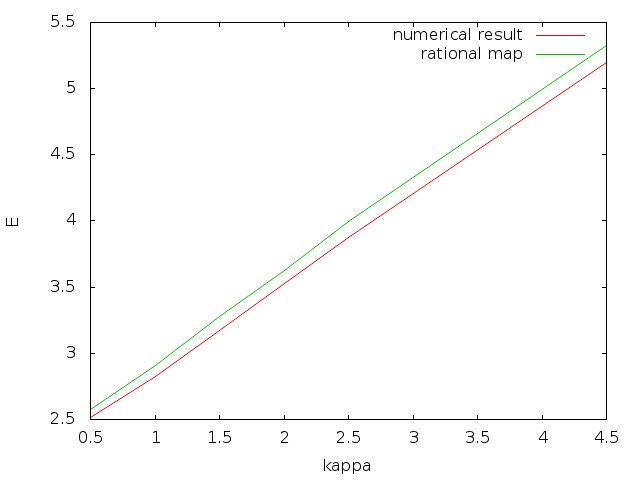}
\end{center}
\caption{The numerical result of the energy compared to the rational map approximation for $B = 2$, for various value of $\kappa$. If you consider the percentage of the approximation that the numerical result takes, it remains roughly constant within our numerical error.}
\label{curvyrational}
\end{figure}

We now compare the approximation from the rational map ansatz to the minimal energy solution for topological charges $B=1$ to $8$. The results for $B=2$ can be observed in figure \ref{curvyrational}. We note that the rational map gives a very good approximation up to $B=4$. The fraction $E_R/E$, where $E_R$ is the energy of the rational map approximation and $E$ is the full numerical minimal energy, seems to stay relatively constant throughout an increase in curvature. We can't say if this trend will definitely continue, however if it does, then the rational maps will remain a good approximation for all values of curvature, as long as the solutions are shell-like, but the rational map approximation breaks down if the solutions begin to become non shell-like.

\begin{table}
\caption{The energy for soliton solutions ($E$) and rational map ansatz  ($E_R$) with $\kappa = 1,m=0$}
\begin{center}
\begin{tabular}{c c c c c c c c}
$B$ & $E$ & $E/B$ & $E_R$ & $E_R/B$ & \% difference & figure \\
\hline
1 & 1.47 & 1.47& 1.47 & 1.47 & 0 & \ref{standardsolutions}(a)\\
2 & 2.82 & 1.41 & 2.90 & 1.45 & 2.9 & \ref{standardsolutions}(b)\\
3 & 4.11 & 1.37 & 4.27 & 1.42 & 3.9 & \ref{standardsolutions}(c)\\
4 & 5.36 & 1.34 & 5.46 & 1.36 & 1.9 & \ref{standardsolutions}(d)\\
5 & 6.66 & 1.33 & 6.89 & 1.38 & 3.4 & \ref{standardsolutions}(e)\\
6 & 7.84 & 1.31 & 8.20 & 1.37 & 4.6 & \ref{standardsolutions}(f)\\
7 & 9.14 & 1.31 & 9.29 & 1.33 & 1.6 & \ref{standardsolutions}(g)\\
8 & 10.29 & 1.29 & 10.73 & 1.34 & 3.9 & \ref{standardsolutions}(h)\\
\hline
\end{tabular}
\end{center}
\label{energies}
\end{table}

\subsection{Crystal chunk Solutions}
For the crystal chunk solutions we will consider a couple of cases, the $B=8$ and $32$ solutions. In Euclidean space we find that the $B=8$ solution needs a relatively high mass for the crystal chunk solution to become the global minima. This massive solution can be considered to be two $B=4$ Skyrmions, joined along an axis perpendicular to a face of the shape. They have relative rotation of $\frac{\pi}{2}$ around the axis joining the two solitons.

\begin{figure}
\begin{center}
\begin{tabular}{c c}
\includegraphics[scale=0.5,natwidth=1000,natheight=1000]{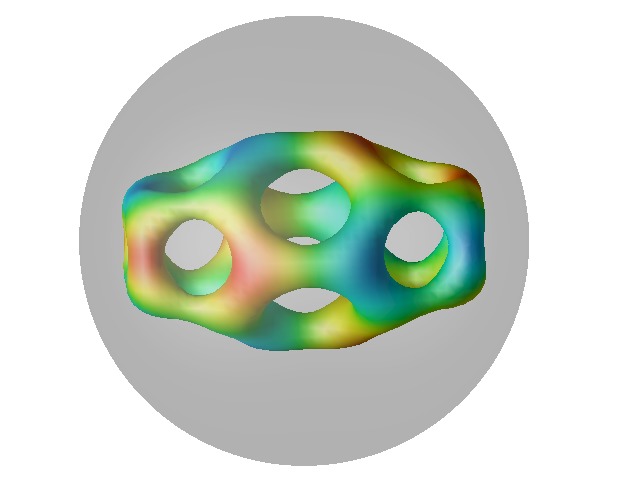} & \includegraphics[scale=0.4,natwidth=1000,natheight=1000]{B8.jpeg} \\
(a) cyrstal-chunk solution & (b) shell-like solution 
\end{tabular}
\end{center}
\caption{$B=8$ static solution, (a) energy density plot of the crystal chunk solution with $\kappa = 1$, $m = 0$, (b) energy density plot of the shell-like solution with $\kappa =1$, $m=0$}
\label{B8}
\end{figure}

In figure \ref{B8} we observe that both the crystal chunk and shell-like solutions are attainable in hyperbolic space with $\kappa = 1$. However, it appears that the crystal solution is the global minima for all non-zero curvatures considered. Note that the energies of the two solutions get very close and could be within numerical error of each other. The $B=8$ crystal chunk solution is the lowest charge crystal solution and hence the energy difference might not be discernible with our accuracy. It is possible that the non-shell like solution does in fact become the minimal energy solution, for higher values of the curvature.  We will consider a higher charge solution where the energy difference will be more discernible. Also, if the crystal chunk solutions act as with increasing the mass term in Euclidean space, we may find that the crystal chunk solution would become the minimal energy solution for a lower curvature.

\begin{figure}
\begin{center}
\begin{tabular}{c c c}
\includegraphics[scale=0.4,natwidth=1000,natheight=1000]{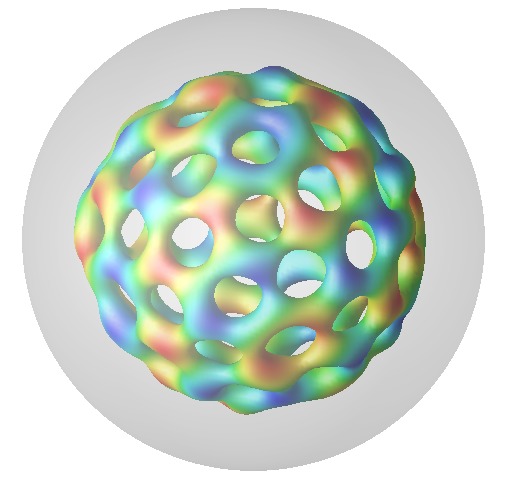} & \includegraphics[scale=0.4,natwidth=1000,natheight=1000]{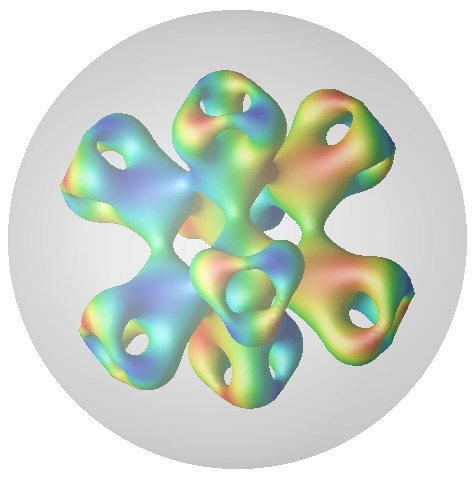} & \includegraphics[scale=0.4,natwidth=1000,natheight=1000]{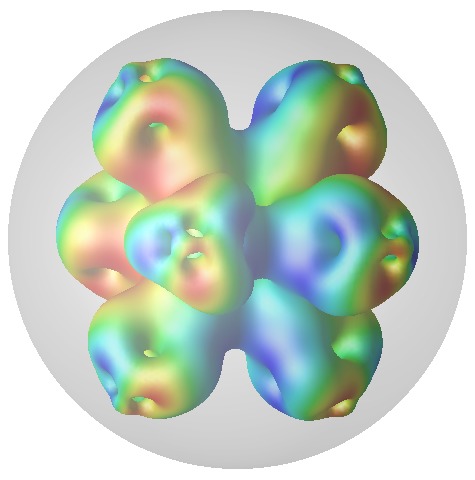}\\
(a) $B=32$ & (b) $B=32$ & (c) $B=32$
\end{tabular}
\caption{Energy density plots of the multi-soliton solution for $B = 32$ for various isosurface values, coloured based on $\pi_2$ value for (a) shell like solution with energy $40.43$, (b-c) crystal chunk solution with energy $38.22$.}
\label{B32}
\end{center}
\end{figure}

The $B=32$ crystal chunk solution, displayed in figure \ref{B32}(b-c), has a far lower energy than that of the shell like solution in figure \ref{B32}(a) for even low values of $\kappa$. This is also the case for a small mass term in the Euclidean model. Hence we have demonstrated that not only are the profile functions related for Skyrmions with massive pions in Euclidean space and with massless pions in hyperbolic space, but the energetically favourable form of solution is also similar. 

\section{Dynamics}
The solutions to the time-dependant equations of motion that follow from (\ref{lagden}) were again found using a time dependent $4$th-order Runga-Kutta method. The grid was modelled using the Poincar\'e ball model of radius $1$ (fixing $\kappa = 1$) on a cubic grid with $(201)^3$ grid point, hence the lattice spacing $\Delta x=0.005$. Spatial derivatives have been approximated using a $4$th-order finite difference method. The product ansatz was used for well separated single charge solitons.

\begin{figure}
  \begin{center}
 \begin{tabular}{c c c c c}
   \includegraphics[width=0.165\linewidth]{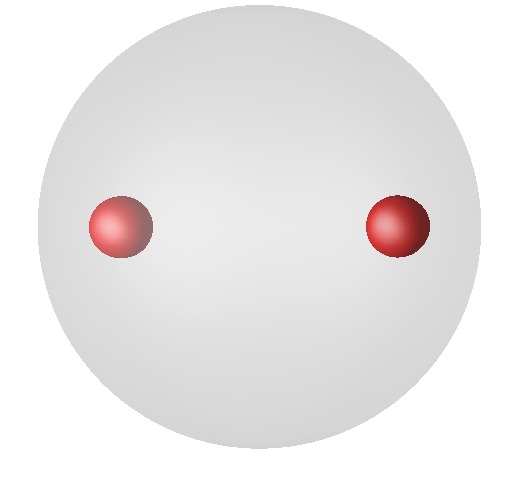}&
   \includegraphics[width=0.165\linewidth]{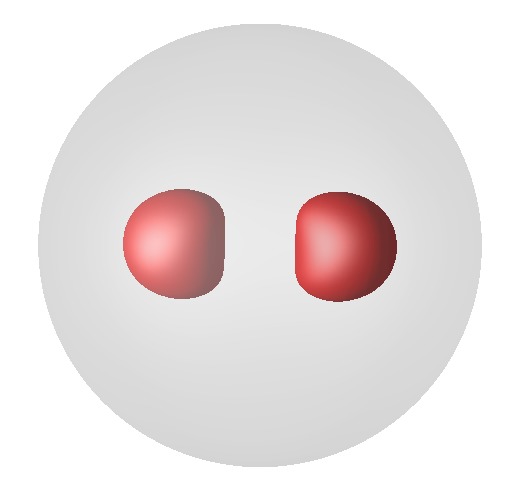}&
   \includegraphics[width=0.165\linewidth]{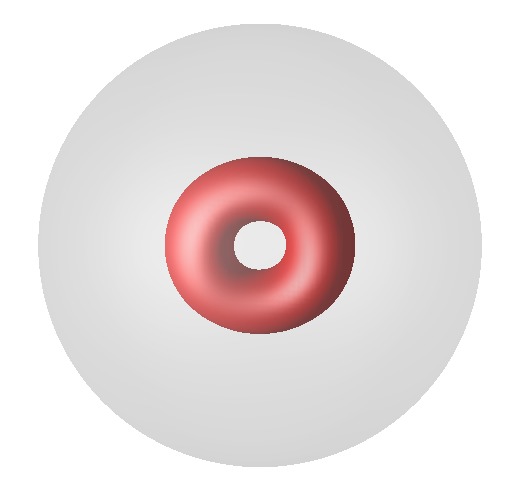}&
   \includegraphics[width=0.165\linewidth]{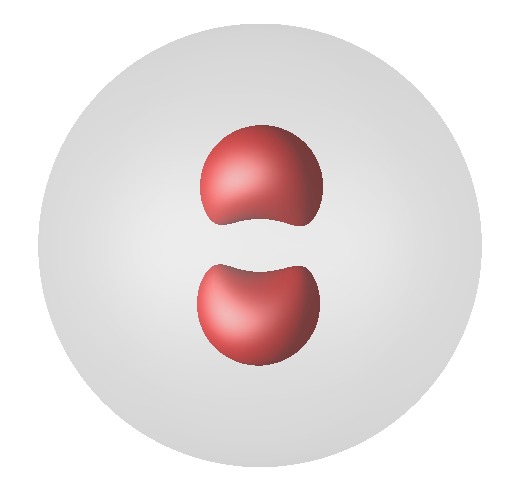}&
   \includegraphics[width=0.165\linewidth]{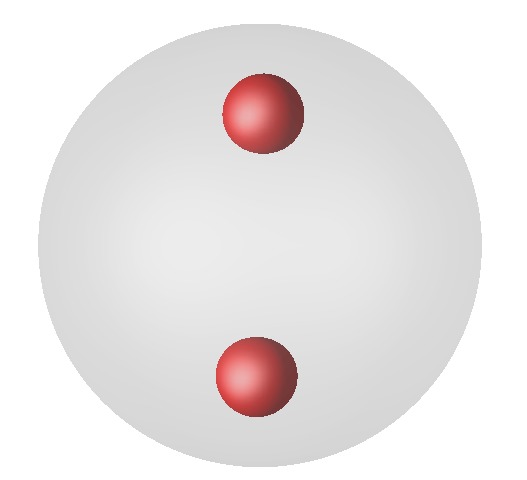}
   \end{tabular}
 \end{center}
 \caption{Scattering along a geodesic through the origin, with zero initial velocity, with solitons in the attractive channel (relative rotation of $\pi$ around a line perpendicular to the diagonal).}
 \label{diagscat}
 \end{figure}

The simplest situation to consider is scattering along a geodesic that passes through the centre of the space, as seen in figure \ref{diagscat}. This gives a straight geodesic, with a clear parallel to Euclidean space and hence an obvious attractive channel (rotate relative by $\pi$ around an axis perpendicular to the connecting straight line). The scattering process then proceeds as expected with the solitons scattering at $\pi/2$. 

\begin{figure}
  \begin{center}
 \begin{tabular}{c c c c c}
   \includegraphics[width=0.155\linewidth]{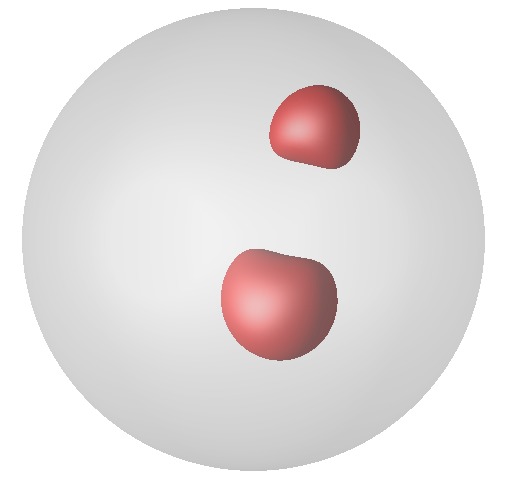}&
   \includegraphics[width=0.155\linewidth]{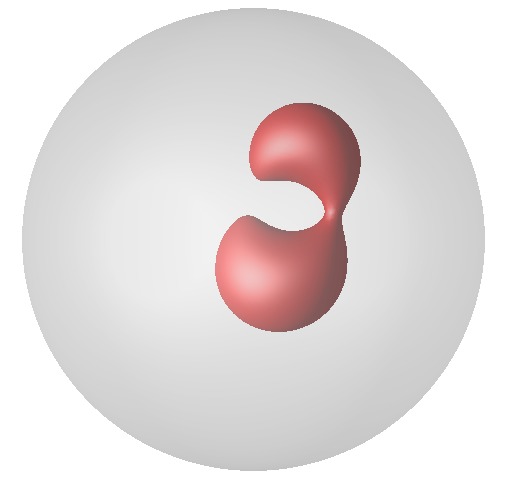}&
   \includegraphics[width=0.155\linewidth]{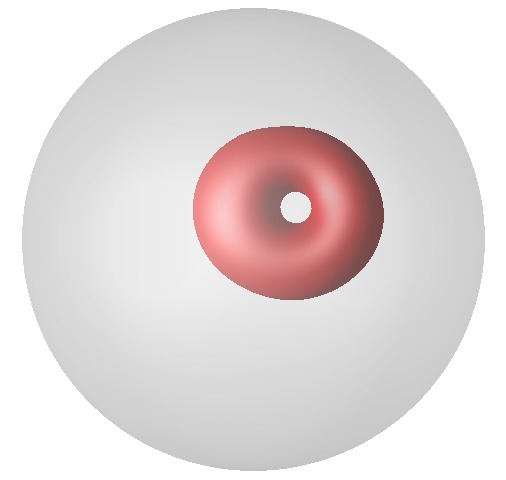}&
   \includegraphics[width=0.155\linewidth]{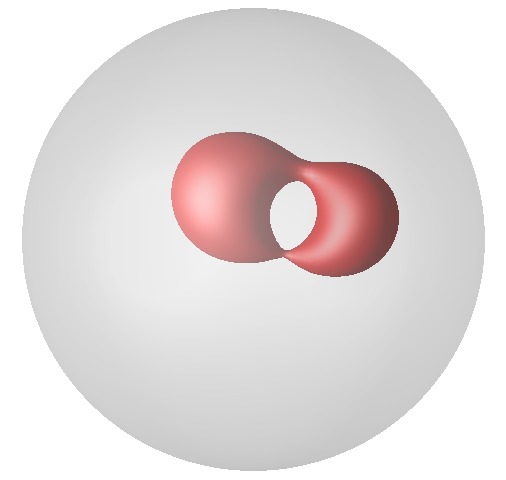}&
   \includegraphics[width=0.155\linewidth]{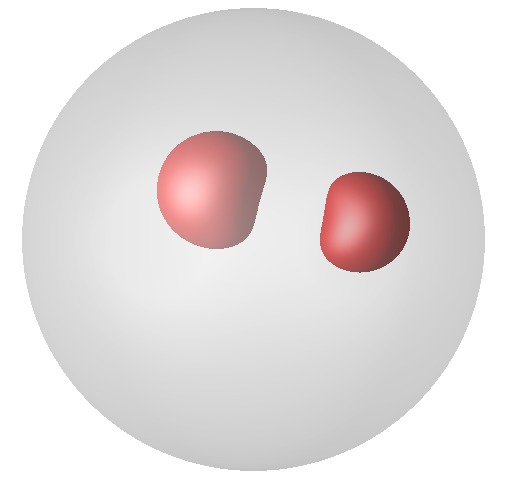}
   \end{tabular}
 \end{center}
 \caption{Scattering along a curved geodesic, with zero initial velocity, in the attractive channel (relative rotation of $\pi$ around a line perpendicular to the geodesic).}
 \label{geoscat}
 \end{figure}

Due to hyperbolic translations (elements of the isometry group of hyperbolic space) one would expect in general, single Skyrmions to follow geodesics until they scatter. After scattering, the emerging Skyrmions will follow alternate geodesics, oriented to the incident paths by a rotation of $\pi$ around an orthogonal axis. The maximal channel will be a rotation of one of the solitons relative to the other by $\pi$ around an orthogonal axis to the tangent of the connecting geodesic. On scattering, the Skyrmions should merge to form the standard $B=2$ solution, oriented to lie in the incident plane, however it may appear deformed due to the curvature of the space. The results presented here confirm these expectations and can be observed in figure \ref{geoscat}.

\section{Conclusions}
We have found both static and dynamic solutions for hyperbolic Skyrmions of various curvature. The static solutions have been related to massive solutions in Euclidean space, by making use of the relation shown in \cite{Atiyah2005106}. It has been demonstrated that the link between curvature in hyperbolic space and mass in Euclidean space extends to full solutions of various topological charge, allowing predictions to be made for the type of solution that will occur in the two models.

We have supplied evidence that suggests the rational map approximation is a good approximation for increasing curvature. It seems to retain its accuracy regardless of the curvature considered. This would suggest that we can model Skyrmion solutions in the infinite curvature limit, by using their respective rational maps. This is analogous to the hyperbolic monopole case, where solutions for infinite curvature become rational maps \cite{BLM:19049}. It would be interesting to see if there were some interesting limit in which it produces exact solutions, that in some way corresponds to hyperbolic monopoles.

The dynamics of various soliton initial conditions have also been studied. The attractive channel was shown to be a relative rotation by $\pi$ around an axis orthogonal to the connecting geodesic.

It would be interesting to consider the form of a soliton crystal in hyperbolic space, due to the interesting symmetries and tilings that can be formed from various polyhedron. It would be sensible to start with the 2-dimensional analogue, due to the difficulty of the task. Some similar work has been done with 2-dimensional vortices in the hyperbolic plane, concentrating on the tiling with Schl{\"a}fi symbol $\left\{8,8\right\}$\cite{Maldonado:2015gfa}.
\section{Acknowledgements}
I would like to thank EPSRC for my PhD studentship. I would also like to thank my supervisor Paul Sutcliffe for useful discussions.

\bibliographystyle{testbib}
\bibliography{hyperbolic_skyrmions}
\end{document}